\documentclass[a4paper,12pt]{article}
\usepackage{amsmath}
\usepackage{amssymb}
\usepackage{amsfonts}
\usepackage{mathrsfs}
\usepackage{cite}
\usepackage[font=small,labelfont=bf]{caption}
\usepackage{graphicx}
\usepackage{lscape}
\usepackage{colordvi}
\usepackage{array}

\makeatletter
\newcommand*\wt[1]{\mathpalette\wthelper{#1}}
\newcommand*\wthelper[2]{%
        \hbox{\dimen@\accentfontxheight#1%
                \accentfontxheight#11.3\dimen@
                $\m@th#1\widetilde{#2}$%
                \accentfontxheight#1\dimen@
        }%
}

\newcommand*\accentfontxheight[1]{%
        \fontdimen5\ifx#1\displaystyle
                \textfont
        \else\ifx#1\textstyle
                \textfont
        \else\ifx#1\scriptstyle
                \scriptfont
        \else
                \scriptscriptfont
        \fi\fi\fi3
}
\makeatother

\def\be{\begin{equation}}
\def\ee{\end{equation}}
\def\beq{\begin{equation}}
\def\eeq{\end{equation}}
\def\bea{\begin{eqnarray}}
\def\eea{\end{eqnarray}}

\def\<{\left\langle}
\def\>{\right\rangle}

\newcommand{\vev}[1]{\left\langle{#1}\right\rangle}
\newcommand{\modulus}[1]{\left|{#1}\right|}

\DeclareMathOperator{\diag}{diag}

\begin{document}

\bibliographystyle{OurBibTeX}

\title{\hfill ~\\[-30mm]
       \hfill\mbox{\small SHEP-12-05}\\[-3mm]
 \hfill\mbox{\small IPPP-12-09}\\[-3mm]
\hfill\mbox{\small DCPT-12-18}\\[13mm]
       \sffamily\Huge{$A_4\times SU(5)$ SUSY GUT of Flavour with Trimaximal Neutrino Mixing}}

\author{
Iain~K.~Cooper$^{1\,}$\footnote{E-mail: \texttt{ikc1g08@soton.ac.uk}}\:,~~
Stephen~F.~King$^{1\,}$\footnote{E-mail: \texttt{king@soton.ac.uk}}~~and~~
Christoph~Luhn$^{2\,}$\footnote{E-mail: \texttt{christoph.luhn@durham.ac.uk}}\\[9mm]
{\small\it
$^1$School of Physics and Astronomy,
University of Southampton,}\\
{\small\it Southampton, SO17 1BJ, U.K.}\\[2mm]
{\small\it
$^2$Institute for Particle Physics Phenomenology,
University of Durham,}\\
{\small\it Durham, DH1 3LE, U.K.}
}

\date{}

\maketitle
\thispagestyle{empty}

\begin{abstract}
Recent T2K, MINOS and Double CHOOZ results, together with global fits of mixing parameters, 
indicate a sizeable reactor
angle~$\theta_{13}\sim 8^\circ$ which, if confirmed, would rule out tri-bimaximal (TB) lepton mixing. 
Recently two of us studied the vacuum alignment of the Altarelli-Feruglio
$A_4$ family symmetry model including additional flavons in the ${\bf 1'}$ and
${\bf 1''}$ representations, leading to so-called ``trimaximal'' 
neutrino mixing and allowing a potentially large reactor angle. Here we
show how such a model may arise from a Supersymmetric (SUSY) 
Grand Unified Theory (GUT) based on
$SU(5)$, leading to sum rule bounds $|s|\leq \frac{\theta_C}{3}$, 
$|a|\leq \frac{1}{2} (r+ \frac{\theta_C}{3})|\cos \delta|$, 
where $s,a,r$ parameterise the solar, atmospheric and reactor
angle deviations from their TB mixing values, 
$\delta$ is the CP violating oscillation phase,
and $\theta_C$ is the Cabibbo angle.

\noindent

\end{abstract}

\newpage
\setcounter{footnote}{0}

\section{Introduction}
Recently T2K have published evidence for a large non-zero reactor angle
\cite{Abe:2011sj}. Combining this with data from MINOS \cite{Adamson:2011qu}
and other experiments in a global fit yields \cite{Schwetz:2011zk}
$\theta_{13}=6.5^\circ\pm 1.5^\circ$, assuming a normal neutrino mass hierarchy.
Here the errors indicate the 1$\sigma$ range, although the statistical
significance of a non-zero reactor angle is about 3$\sigma$. 
Other global fits indicate a larger reactor angle $\theta_{13}=9.1^\circ \pm
1.3^\circ $ \cite{Fogli:2011qn}.\footnote{In both cases, the quoted ranges are
calculated using the new reactor anti-neutrino fluxes \cite{Mueller:2011nm}.}
The first results from Double CHOOZ also show indications of a non-zero
reactor angle \cite{Abe:2011fz}. If confirmed, a sizable reactor angle
$\theta_{13}\sim 8^\circ$, consistent with both global fits,
would rule out the hypothesis of exact
tri-bimaximal (TB) mixing~\cite{Harrison:2002er}. 

In the framework of Supersymmetric (SUSY) Grand Unified Theories (GUTs)
of Flavour \cite{Gutflav} (i.e. with a Family Symmetry \cite{Reviews}
implemented) it is already known that TB mixing cannot be exact.
The reason is that, even if
TB mixing is realised exactly in the neutrino sector, observable lepton mixing is subject to
charged lepton (CL) corrections
(due to the fact that $U_{\mathrm{PMNS}}=U_{e}U_{\nu}^{ \dagger}$)
and renormalisation group (RG) corrections, not
to mention other corrections due to canonical normalisation (CN)
(for a unified discussion of all three corrections
see e.g. \cite{Antusch:2008yc} and references therein). Therefore,
in the framework of SUSY GUTs of Flavour, the question of whether TB mixing
may be maintained in the neutrino sector is a quantitative one: can the above
CL, RG and CN corrections be sufficiently large to account for the observed
reactor angle? The answer is yes in some cases (see
e.g. \cite{Antusch:2011qg}), but no in many other cases.  For example, in
models based on the Georgi-Jarlskog mechanism \cite{Georgi:1979df}, where the
CL corrections are less than or about $3^\circ$, and where the RG and CN corrections are less than or about
$1^\circ$ (which is the case for
hierarchical neutrinos), it would be difficult to account for a reactor angle
$\theta_{13}\sim 8^\circ$.  For this reason, there is a good motivation to consider other
patterns of neutrino mixing beyond TB mixing, and many alternative proposals
\cite{flurry} have indeed been put forward to account for a non-zero $\theta_{13}$. 
On the other hand, since the solar and atmospheric mixing angles remain
consistent with TB mixing, there is also a good motivation to maintain these successful
predictions of TB mixing. 

In going beyond TB mixing, it is useful to consider a general parameterisation of the 
Pontecorvo-Maki-Nakagawa-Sakata (PMNS) mixing matrix in terms of deviations
from TB mixing \cite{King:2007pr},  
\begin{eqnarray}
U_{\mathrm{PMNS}} \approx
\left( \begin{array}{ccc}
{\frac{2}{\sqrt{6}}}(1-\frac{1}{2}s)  & \frac{1}{\sqrt{3}}(1+s) & \frac{1}{\sqrt{2}}re^{-i\delta } \\
-\frac{1}{\sqrt{6}}(1+s-a + re^{i\delta })  & \phantom{-}\frac{1}{\sqrt{3}}(1-\frac{1}{2}s-a- \frac{1}{2}re^{i\delta })
& \frac{1}{\sqrt{2}}(1+a) \\
\phantom{-}\frac{1}{\sqrt{6}}(1+s+a- re^{i\delta })  & -\frac{1}{\sqrt{3}}(1-\frac{1}{2}s+a+ \frac{1}{2}re^{i\delta })
 & \frac{1}{\sqrt{2}}(1-a)
\end{array}
\right)P\ ,~~
\label{PMNS1}
\end{eqnarray}
where the deviation parameters $s,a,r$ are defined as \cite{King:2007pr}, 
\begin{eqnarray}
&&
 \sin \theta_{12} = \frac{1}{\sqrt{3}}(1+s)\ , \qquad
\sin \theta_{23}  =  \frac{1}{\sqrt{2}}(1+a)\  , \qquad 
 \sin \theta_{13} =   \frac{r}{\sqrt{2}}\ .
\label{rsa}
\end{eqnarray}
For example, the global fit of the conventional mixing angles in
\cite{Schwetz:2011zk} can be translated into the $1\sigma$ ranges:
\begin{eqnarray}
&& 0.12<r<0.20,\ \ -0.06<s<-0.006, \ \  -0.06<a<0.08.
\label{rsafit}
\end{eqnarray}

In a SUSY GUT of Flavour, the Family Symmetry is responsible for determining
the neutrino mixing pattern, which then gets corrected by CL, RG and CN
contributions to yield the observed lepton mixing angles. The question is what
is the underlying neutrino mixing pattern? If we want to go beyond TB neutrino
mixing, there are many possibilities. One simple scheme is
the trimaximal (TM) mixing pattern \cite{Haba:2006dz}:
\begin{equation}
\label{TM}
U^{\nu \dagger}_{\mathrm{TM}}~=~P'\begin{pmatrix} 
\frac{2}{\sqrt{6}}\cos\vartheta
&\frac{1}{\sqrt{3}}
&\frac{2}{\sqrt{6}}\sin\vartheta \,e^{i\rho}\\
-\frac{1}{\sqrt{6}}\cos\vartheta -\frac{1}{\sqrt{2}}\sin\vartheta\,e^{-i\rho}
&\frac{1}{\sqrt{3}}
& \phantom{-}\frac{1}{\sqrt{2}}\cos\vartheta-\frac{1}{\sqrt{6}}\sin\vartheta  \,e^{i\rho} \\ 
-\frac{1}{\sqrt{6}}\cos\vartheta+ \frac{1}{\sqrt{2}}\sin\vartheta\,e^{-i\rho}
& \frac{1}{\sqrt{3}}
& - \frac{1}{\sqrt{2}}\cos\vartheta-\frac{1}{\sqrt{6}}\sin\vartheta \,e^{i\rho} 
\end{pmatrix} P \ ,
\end{equation}
where $\frac{2}{\sqrt{6}} \sin \vartheta=\sin\theta_{13}^{\nu}$, $P'$ is a
diagonal phase matrix required to put 
$U_{\mathrm{PMNS}}=U^{e}U^{\nu \dagger}_{\mathrm{TM}}$
into the PDG
convention \cite{Nakamura:2010zzi}, and $P={\rm
  diag}(1,e^{i\frac{\alpha_{2}}{2} }, e^{i\frac{\alpha_{3}}{2} })$ contains
the usual Majorana phases. In particular TM mixing approximately predicts TB
neutrino mixing for the solar neutrino mixing angle $\theta_{12}^{\nu}\approx
35^\circ$ as the correction due to a non-zero but relatively small reactor
angle is of second order. However we emphasise again that, in a SUSY GUT of
Flavour, TM mixing refers to the neutrino mixing angles only, and the physical
lepton mixing angles will involve additional CL, RG and CN
corrections. Nevertheless, TM neutrino mixing could provide a better starting
point than TB neutrino mixing, if the reactor angle proves to be $\theta_{13}\sim 8^\circ$, 
and this provides the motivation for the approach
followed in this paper. 

Recently, an $A_4$ model of TM neutrino mixing was discussed in \cite{King:2011zj}.
In the original $A_4$ models of TB mixing Higgs fields \cite{Ma:2001dn} or
flavon fields \cite{Altarelli:2005yx} transforming under $A_4$ as ${\bf 3}$
and ${\bf 1}$ but not ${\bf 1'}$ or ${\bf 1''}$ were used to break the family
symmetry and to lead to TB mixing. However there is no good reason not to
include flavons transforming as ${\bf 1'}$ or ${\bf 1''}$, and once included
they will lead to deviations from TB mixing \cite{Brahmachari:2008fn}, 
in particular it was noted that they lead to TM mixing \cite{Shimizu:2011xg}.
In \cite{King:2011zj} the vacuum alignment of the Altarelli-Feruglio $A_4$
family symmetry model \cite{Altarelli:2005yx}, including additional flavons in
the ${\bf 1'}$ and/or ${\bf 1''}$ representations, was studied and it was shown
that it leads to TM neutrino mixing. In this paper we shall show how such a
model may arise from a SUSY GUT based on $SU(5)$, leading to the sum rule bounds
$|s|\leq \frac{\theta_C}{3}$ and 
$|a|\leq \frac{1}{2} (r+ \frac{\theta_C}{3}) |\cos \delta|$, up to RG and CN corrections,
where $r,s,a$ are the above tri-bimaximal deviation parameters, 
$\delta$ is the CP violating oscillation phase,
and $\theta_C$ is the Cabibbo angle. 
However we shall not be interested in the details of the GUT breaking Higgs potential,
which is model dependent and may not even be describable by renormalisable
4d field theory, as is the case where 
GUT breaking is due to orbifold constructions \cite{Burrows:2009pi}. Instead we shall take the
GUT breaking as ``given'' and formulate the theory just below the GUT scale,
although we shall use $SU(5)$ notation for convenience, and to emphasise the
phenomenological predictions which follow from GUTs. 

The rest of the paper is organised as follows. In Section \ref{sec:model} we
introduce the model, presenting field content, charges, flavon alignments and
leading order (LO) superpotential terms. Section \ref{sec:masses} then
presents the mass matrices and mixing angles for neutrinos, quarks and charged
leptons arising from the LO superpotential. The effect of the non-trivial charged lepton
corrections (due to the grand unified setup) on the physical lepton mixing
angles is discussed in Section \ref{sec:pmns}. 
We conclude in Section \ref{sec:conclusion}. The discussion of the vacuum
alignment and the next to leading order (NLO) terms is presented in
Appendices \ref{sec:appenda} and \ref{sec:appendb}, respectively.  


\section{The model}

\begin{table}
	\centering
		\begin{tabular}{|c||c|c|c|c|c||c|c|c|}
			\hline
				Field & $N$ & $F$ & $T_1$ & $T_2$ & $T_3$ & $H_{\bf 5}$ & $H_{\bf \overline{5}}$ & $H_{\bf \overline{45}}$ \\ \hline \hline
				$\!SU(5)\!$ & ${\bf 1}$ & ${\bf \overline{5}}$ & ${\bf 10}$ & ${\bf 10}$ & ${\bf 10}$ & ${\bf 5}$ & ${\bf \overline{5}}$ & ${\bf \overline{45}}$ \\ \hline
				$A_4$ & ${\bf 3}$ & ${\bf 3}$ & ${\bf 1''}$ & ${\bf 1'}$ & ${\bf 1}$ & ${\bf 1}$ & ${\bf 1'}$ & ${\bf 1''}$ \\ \hline
				$\!{U}(1)_R\!$ & $1$ & $1$ & $1$ & $1$ & $1$ & $0$ & $0$ & $0$ \\ \hline\hline
				$U(1)$ & $1$ & $-1$ & $3$ & $3$ & $0$ & $0$ & $-1$ & $-2$ \\ \hline
				$Z_2$ & $+$ & $+$ & $+$ & $+$ & $+$ & $+$ & $+$ & $-$ \\ \hline
                                $Z_3$ & $\omega$ & $\omega^2$ & $\omega^2$ & $1$ & $1$ & $1$ & $\omega$ & $\omega$ \\ \hline
				$Z_5$ & $\rho$ & $\rho^4$ & $1$ & $1$ & $1$ & $1$ & $\rho$ & $\rho$ \\ \hline
		\end{tabular}
	\caption{Matter and Higgs chiral superfields in the model. }
\label{tab:matter}
\end{table}

The transformation properties of the $SU(5)$ matter and Higgs multiplets are
shown in Table~\ref{tab:matter}. $N$ denotes the right-handed neutrino fields,
$F$ the ${\bf{\overline{5}}}$ of $SU(5)$, containing the lepton doublet and
the right-handed down-type quark, and $T$ labels the ${\bf{10}}$ which
includes the quark doublet as well as the right-handed up-type quark and
charged lepton. $N$ and $F$ furnish the triplet representation of $A_4$, thus
unifying the three families of leptons, while the three families of the $T_i$
transform in the three distinct one-dimensional representations of $A_4$.
In the Higgs sector, we have introduced the $H_{\bf{\overline{45}}}$ in order
to implement the Georgi-Jarlskog mechanism \cite{Georgi:1979df}.\footnote{The standard
MSSM $\mu$-term $\mu H_u H_d$ (where $H_u$ is the $SU(2)_L$ doublet of $H_{\bf 5}$;
  and $H_d$ is a linear combination of the $SU(2)_L$ doublets in $H_{{\bf
      \overline{5}}}$ and~$H_{{\bf \overline{45}}}$) is
forbidden by the $A_4$, $U(1)$, $Z_3$ and $Z_5$ symmetries as well as
$U(1)_R$, allowing for a natural solution to the $\mu$-problem of the MSSM
using a GUT singlet from the hidden sector of Supergravity theories
\cite{Giudice:1988mu}.}

The full set of flavon fields is shown in Table~\ref{tab:flavons}. The fields
$\varphi_S$ and $\xi^i$ are responsible for the flavour structure of the neutrino
sector, while the flavons $\varphi_T$ and $\theta^i$ control the quark and
charged lepton sector. The vacuum structure is obtained via the standard
$F$-term alignment mechanism \cite{Altarelli:2005yx} where the $F$-terms of
so-called driving fields (denoted by a superscript $0$) are set to zero, thus
giving rise to constraints which in turn fix the flavon alignments. As shown in
Appendix~\ref{sec:appenda}, one obtains the following triplet flavon
alignments,\footnote{The auxiliary flavon field $\sigma$ is introduced for the purpose
  of achieving the alignment of the $U(1)$ charged flavon field $\varphi_T$.}
\be
\left\langle\varphi_T\right\rangle  ~\propto~ \begin{pmatrix} 1\\0\\0 \end{pmatrix},
\qquad
\left\langle\varphi_S\right\rangle  ~\propto~ \begin{pmatrix}
  1\\1\\1 \end{pmatrix} . \label{eq:flavonalign}
\ee

The $U(1)_R$ represents an $R$-symmetry whose $Z_2$ subgroup gives rise to the
standard $R$-parity which forbids unwanted operators contributing to
proton decay and keeps the lightest SUSY particle a good candidate for cold
dark matter. The $U(1)$ and the three $Z_N$ shaping symmetries constrain the
structure of the Yukawa matrices in the quark and charged lepton
sectors. Specifically, the $Z_5$ prevents the neutrino flavons ($\varphi_S$
and $\xi^i$) from appearing in the quark and charged lepton Yukawa couplings.

\begin{table}
	\centering
		\begin{tabular}{|c||c|c|c|c||c|c|c|c|c|c||c|c|c|}
			\hline
				Field & $\varphi^{\vphantom{0}}_{S}$ & $\xi\vphantom{\Big(}$  & $\xi'$ & $\xi''$ & $\varphi^{\vphantom{0}}_T$ & $\theta$ & $\theta'$ & $\theta''$ & $\wt{\theta}'$ & $\sigma$ & $\varphi_T^0$ & $\varphi_S^0$ & $\xi^0$ \\ \hline \hline
				$\!SU(5)\!$ & ${\bf 1}$ & ${\bf 1}$ & ${\bf
                                  1}$ & ${\bf 1}$ & ${\bf 1}$ & ${\bf 1}$ &
                                ${\bf 1}$ & ${\bf 1}$ & ${\bf 1}$ & ${\bf 1}$
                                & ${\bf 1}$ & ${\bf 1}$ & ${\bf 1}$ \\ \hline
				$A_4$ & ${\bf 3}$ & ${\bf 1}$ & ${\bf 1'}$ & ${\bf 1''}$ & ${\bf 3}$ & ${\bf 1}$& ${\bf 1'}$ & ${\bf 1''}$ & ${\bf 1'}$ & ${\bf 1}$ & ${\bf 3}$ & ${} \bf 3$ & ${\bf 1}$ \\ \hline
				$\!{U}(1)_R\!$ & $0$ & $0$ & $0$ & $0$ & $0$ & $0$ & $0$ & $0$ & $0$ & $0$ & $2$ & $2$ & $2$\\ \hline\hline
				$U(1)$ & $-2$ & $-2$ & $-2$ & $-2$ & $2$ & $-1$ & $-1$ & $-1$ & $-5$ & $2$ & $-4$ & $4$ & $4$ \\ \hline
				$Z_2$ & $+$ & $+$ & $+$ & $+$ & $+$ & $-$ & $+$ & $+$ &$-$ & $+$ & $+$ & $+$ & $+$ \\ \hline
                                $Z_3$ & $\omega$ & $\omega$ & $\omega$ & $\omega$ & $1$ & $\omega$ & $\omega^2$ & $\omega^2$ & $\omega^2$ & $1$ & $1$ & $\omega$ & $\omega$ \\ \hline
				$Z_5$ & $\rho^3$ & $\rho^3$ & $\rho^3$ & $\rho^3$ & $1$ & $1$ & $1$ & $1$ & $1$ & $1$ & $1$ & $\rho^4$ & $\rho^4$ \\ \hline
		\end{tabular}
	\caption{Flavon chiral superfields in the model. }
\label{tab:flavons}
\end{table}

In the neutrino sector, the $A_4$ family symmetry is broken by the flavon
fields $\varphi_S$ and $\xi^i$, thereby leading to a TM mixing pattern as
observed in \cite{King:2011zj}.  In the quark and charged lepton sector the
$A_4$ symmetry is broken differently by virtue of the flavon fields $\varphi_T$ and
$\theta^i$. Due to the $SU(5)$ structure, the form of the charged lepton and
down quark Yukawa matrices is intimately related, leading to a
non-trivial left-handed charged lepton mixing which combines with the TM
structure of the neutrino mixing to give the physical PMNS mixing.

\label{sec:model}

\subsection{Allowed terms}

The neutrino sector is composed of Dirac and Majorana mass terms which take
the leading order form in the superpotential,
\begin{equation}
{W}_{\nu}=yFNH_{\bf 5} +
\left(y_1\varphi_S+y_2\xi+y_2'\xi'+y_2''\xi''\right)NN \ ,
\label{eqn:neutrino}
\end{equation}
with  $y,\;y_1,\;y_2,\;y_2',\;y_2''$ being dimensionless couplings.

The leading order superpotential terms of the down quark and charged lepton
sector are given as follows
\begin{equation}
\begin{split}
{W}_d&~\sim~\left(\frac{\theta^2\theta''}{\Lambda_d^4}\left(F\varphi_T\right)'+\frac{\theta^2\theta'}{\Lambda_d^4}\left(F\varphi_T\right)''\right)H_{\bf \overline{5}}T_1+\frac{\sigma\theta\theta'\left(\theta''\right)^2}{\Lambda_d^6}\left(F\varphi_T\right)H_{\bf\overline{45}}T_1\\
&~~~+~\frac{\left(\theta'\right)^2\theta''}{\Lambda_d^4}\left(F\varphi_T\right)H_{\bf \overline{5}}T_2
+\left(\frac{\theta\theta''}{\Lambda_d^3}\left(F\varphi_T\right)'+\frac{\theta\theta'}{\Lambda_d^3}\left(F\varphi_T\right)''\right)H_{\bf \overline{45}}T_2\\
&~~~+~\left(\frac{\sigma ^2 \theta^2\left(\theta '\right)^2
}{\Lambda_d^7}(F\varphi_T)+\frac{1}{\Lambda_d}\left(\left(F\varphi_T\right)''\right)\right)H_{\bf
  \overline{5}}T_3+\left(\frac{\sigma ^2 \theta^3
}{\Lambda_d^6}(F\varphi_T)'\right)H_{\bf \overline{45}}T_3 \ ,
\label{eqn:down}
\end{split}
\end{equation}
where $\Lambda_d$ is the relevant messenger mass. Note that for some entries
of the down quark Yukawa matrix, there are several different operators of the
same order; here we simply choose an example for illustrative purposes. The
flavons $\theta^i$ play a role similar to a Froggatt-Nielsen field
\cite{Froggatt:1978nt}.  

Finally the leading order up quark sector Yukawa superpotential terms
take the form 
\begin{equation}
\begin{split}
{W}_u&~\sim~\frac{\theta^4\left(\theta'\right)^2}{\Lambda_u^6}T_1T_1H_{\bf 5}+\left(\frac{\theta^2\left(\theta'\right)^2\left(\theta''\right)^2}{\Lambda_u^6}+\frac{\sigma\theta\left(\theta'\right)^2\widetilde{\theta}'}{\Lambda_u^5}\right)(T_1T_2+T_2T_1)H_{\bf 5} \\
&~~~+~\frac{\theta^2\theta'}{\Lambda_u^3}(T_1T_3+T_3T_1)H_{\bf 5} 
+\frac{\theta\widetilde{\theta}'}{\Lambda_u^2}T_2T_2H_{\bf 5}+\frac{\theta'\left(\theta''\right)^2}{\Lambda_u^3}(T_2T_3+T_3T_2)H_{\bf 5}+T_3T_3H_{\bf 5}.
\end{split}
\label{eqn:up}
\end{equation}
It should be mentioned that the messenger mass in this sector, $\Lambda_u$,
may in principle be different from that in the down quark sector. The field
$\widetilde{\theta}'$ is introduced specifically to generate the $T_2T_2$ term to the
required order.

Examples of the many subleading higher order operators allowed by the
symmetries of the model are listed in Appendix \ref{sec:appendb}. As their
contribution to the mass matrices is negligible, they do not induce physically
relevant modifications of the LO picture.

\section{Fermion mass matrices}
\label{sec:masses}
After spontaneous breakdown of the $A_4$ family symmetry by the flavon VEVs,
the superpotential terms of Eqs.~\eqref{eqn:neutrino}-\eqref{eqn:up} predict
mass matrices for the respective sectors. In the 
following, order one coefficients in the quark and charged lepton sectors are
omitted (including flavon VEV normalisation factors). Regarding the scale of
the flavon VEVs we define
\begin{equation}
\eta_i = \frac{\left\langle|\varphi_i|\right\rangle}{\Lambda} ,
\label{eqn:eta}
\end{equation}
where $\varphi_i$=$\varphi_{T}$, $\theta^i$ or $\sigma$. In order to get the
hierarchical structure of the quark and charged lepton mass matrices we assume 
\begin{equation}
\eta_{\widetilde{\theta'}}=\epsilon^2\;
~~\mathrm{and}\;~~ \eta_{\mathrm{others}}=\epsilon ,
\label{eqn:epsilon}
\end{equation}
where the numerical values for $\epsilon$ depend on the messenger scale of the
relevant sector. We present LO operators for each entry in the mass matrices; NLO operators can be found in Appendix \ref{sec:appendb}.

In the Higgs sector, it is not the $H_{\bf 5}$, $H_{{\bf \overline{5}}}$ or
$H_{{\bf \overline{45}}}$ which get VEVs but their $SU(2)_L$ doublet
components. These are the two MSSM doublets $H_u$ (corresponding to
$H_{\bf 5}$) and $H_d$ (corresponding to a linear combination of
$H_{{\bf \overline{5}}}$ and $H_{{\bf \overline{45}}}$); they originate below the
GUT scale and remain massless down to the electroweak scale. The non-MSSM
states all acquire GUT scale masses, including the linear combination of
$H_{{\bf \overline{5}}}$ and $H_{{\bf \overline{45}}}$ orthogonal to
$H_d$. Electroweak symmetry is broken after the light MSSM doublets $H_{u,d}$
acquire VEVs $v_{u,d}$ and they then generate the fermion masses.

In the following all quark and charged lepton mass matrices are given in the
L-R convention, i.e. the mass term for a field $\psi$ is given in the order
$\overline{\psi_L}  M_{LR}\psi_R$.

\subsection{Neutrino sector}
Eq.~\eqref{eqn:neutrino} gives Dirac and Majorana mass matrices
\begin{equation}
 \label{eqn:Dirnumasses}
m_D=\begin{pmatrix}
     1 & 0 & 0 \\
     0 & 0 & 1 \\
     0 & 1 & 0
    \end{pmatrix}yv_u \ ,
\end{equation}
and
\begin{equation}
\label{eqn:Majnumasses}
M_{R}=\left[\alpha\begin{pmatrix} 
	2 & -1 & -1 \\
	-1 & 2 & -1 \\
	-1 & -1 & 2
      \end{pmatrix}+\beta\begin{pmatrix} 
	1 & 0 & 0 \\
	0 & 0 & 1 \\
	0 & 1 & 0
      \end{pmatrix}+\gamma'\begin{pmatrix} 
	0 & 0 & 1 \\
	0 & 1 & 0 \\
	1 & 0 & 0
      \end{pmatrix}+\gamma''\begin{pmatrix} 
	0 & 1 & 0 \\
	1 & 0 & 0 \\
	0 & 0 & 1
      \end{pmatrix}\right],
\end{equation}
with $\alpha=y_1\vev{\varphi_S}$, $\beta=y_2\vev{\xi}$,
$\gamma'=y_2'\vev{\xi'}$ and $\gamma''=y_2''\vev{\xi''}$. As shown in
\cite{King:2011zj}, the standard type I seesaw  formula then yields a light
neutrino mass matrix of trimaximal structure, and hence a neutrino mixing
matrix of the form as given in Eq.~\eqref{TM}.

\subsection{Down quark and charged lepton sector}
In the down quark and charged lepton sector, the superpotential of
Eq.~\eqref{eqn:down} predicts a mass matrix of the form (with messenger mass
$\Lambda_d$ in $\eta_i$)
\begin{equation}
 \begin{pmatrix}
      k_f\eta_{\sigma}\eta_{\theta}\eta_{\theta'}\eta_{\theta''}^2 & \eta_{\theta}^2\eta_{\theta''} & \eta_{\theta}^2\eta_{\theta'} \\
      \eta_{\theta'}^2\eta_{\theta''} & k_f\eta_{\theta}\eta_{\theta''} & k_f\eta_{\theta}\eta_{\theta'} \\
      \eta_{\sigma}^2\eta_{\theta}^2\eta_{\theta'}^2 & k_f\eta_{\sigma}^2\eta_{\theta}^3 & 1
     \end{pmatrix}\eta_{T}v_d \ ,
\label{eqn:downmass1}
\end{equation}
where this matrix has to be transposed for the charged leptons. $k_f$ is the
Georgi-Jarlskog factor which takes the values
\begin{displaymath}
 k_f=\begin{cases}
      1 &\text{for} \quad f=d, \\
      -3 &\text{for} \quad f=e.
     \end{cases}
\end{displaymath}
Inserting the $\epsilon$ suppressions of the flavon VEVs from
Eq.~(\ref{eqn:epsilon})  the down quark mass matrix becomes
\begin{equation}
 M_d\sim\begin{pmatrix}
         \epsilon^5 & \epsilon^3 & \epsilon^3 \\
	 \epsilon^3 & \epsilon^2 & \epsilon^2 \\
	 \epsilon^6 & \epsilon^5 & 1
        \end{pmatrix}\epsilon \,v_d
\label{eqn:downmass2},
\end{equation}
whilst the charged lepton mass matrix reads
\begin{equation}
 M_e\sim\begin{pmatrix}
         -3\epsilon^5 & \epsilon^3 & \epsilon^6 \\
	 \epsilon^3 & -3\epsilon^2 & -3\epsilon^5 \\
	 \epsilon^3 & -3\epsilon^2 & 1
        \end{pmatrix}\epsilon\, v_d
\label{eqn:chargedmass}.
\end{equation}
Here we assume the numerical value $\epsilon\sim0.15$.
Upon diagonalisation, these give mass ratios of $\epsilon^4:\epsilon^2:1$ for
the down-type quarks and $\frac{\epsilon^4}{3}:3\epsilon^2:1$ for the charged
leptons. These ratios are in good agreement with quark and lepton data and
also predict the Georgi-Jarlskog GUT scale mass relations of $m_e \sim
\frac{m_d}{3}$, $m_{\mu}\sim 3m_{s}$ and $m_{\tau}\sim m_b$ as desired. In the
low quark angle approximation, the left-handed down quark mixing angles
$\theta^d_{12}\sim\epsilon$, $\theta^d_{13}\sim\epsilon^3$ and
$\theta^d_{23}\sim\epsilon^2$ are also predicted in agreement with data
(assuming an approximately diagonal up quark sector which we obtain in the next
subsection). The corresponding charged lepton mixing angles are
$\theta^e_{12}\sim\frac{\epsilon}{3}$, $\theta^e_{13}\sim\epsilon^6$ and
$\theta^e_{23}\sim3\epsilon^5$. Therefore, the only significant charged lepton
correction to the TM mixing of the neutrino sector originates from
$\theta^e_{12}\sim \frac{\theta_C}{3}$, where $\theta_C$ denotes the Cabibbo angle.

\subsection{Up quark sector}
Eq.~\eqref{eqn:up} may be expanded after $A_4$
symmetry breaking and is responsible for up-type quark masses
\begin{equation}
 \begin{pmatrix}
      \eta_{\theta}^4\eta_{\theta'}^2 & \eta_{\theta}^2\eta_{\theta'}^2\eta_{\theta''}^2 + \eta_{\sigma}\eta_{\theta}\eta_{\theta'}^2\eta_{\widetilde{\theta}'}
 & \eta_{\theta}^2\eta_{\theta'} \\
      \eta_{\theta}^2\eta_{\theta'}^2\eta_{\theta''}^2 + \eta_{\sigma}\eta_{\theta}\eta_{\theta'}^2\eta_{\widetilde{\theta}'} & \eta_{\theta}\eta_{\widetilde{\theta'}} & \eta_{\theta'}\eta_{\theta''}^2 \\
      \eta_{\theta}^2\eta_{\theta'} & \eta_{\theta'}\eta_{\theta''}^2 & 1
     \end{pmatrix}v_u\ .
\label{eqn:upmass1}
\end{equation}
Taking the VEV hierarchy as in Eq.~\eqref{eqn:epsilon}, but now adopting the
messenger scale $\Lambda_u \approx \frac{3}{2}\Lambda_d$, we obtain a mass
matrix with an expansion parameter $\overline{\epsilon}\sim0.1$,
\begin{equation}
 M_u\sim\begin{pmatrix}
      \overline{\epsilon}^6 & \overline{\epsilon}^6 & \overline{\epsilon}^3 \\
      \overline{\epsilon}^6 & \overline{\epsilon}^3 & \overline{\epsilon}^3 \\
      \overline{\epsilon}^3 & \overline{\epsilon}^3 & 1
     \end{pmatrix}v_u \ .
\label{eqn:upmass2}
\end{equation}
and an up-type quark mass hierarchy
$\overline{\epsilon}^6:\overline{\epsilon}^3:1$. This matrix gives mixing
angles of
$\theta^u_{12}\sim\theta^u_{13}\sim\theta^u_{23}\sim\overline{\epsilon}^3$. This
means that the Cabibbo-Kobayashi-Maskawa (CKM) mixing matrix is dominated by
down quark mixing, except that there may be a contribution to
$\theta^{\mathrm{CKM}}_{13}$ from the up quark sector which is almost as significant
as the contribution coming from the down-type quarks. The Cabibbo angle is still
approximately $\theta_C \sim \theta_{12}^d \sim \epsilon$.

\section{ Charged lepton corrections to lepton mixing}
\label{sec:pmns}
We have presented mixing angles which rotate the charged leptons and neutrino
fields between the mass and flavour bases, however these individual rotations
are not what experiments observe. It is the combination of the two mixing matrices
that appears in the electroweak coupling to the $W$ boson, giving the physical
mixing matrix
\begin{equation}
 U_{\mathrm{PMNS}}=U_{e_L}U_{\nu_L}^{\dagger}.
\label{eqn:PMNS}
\end{equation}
Here it is understood that for Lagrangians in the left right convention,
$U_{e_L}$ acts as $U_{e_L}m^{}_e m_e^\dagger
U_{e_L}^{\dagger}=(m_e^{\mathrm{diag}})^2$ and $U_{\nu_L}$ as
$U_{\nu_L}m_{\nu}U_{\nu_L}^T=m_{\nu}^{\diag}$.  
While the neutrino sector predicts exact TM mixing, the effect of the charged
lepton corrections generates an experimentally detectable deviation from
this in the physical parameters. In this section we ignore RG and CN corrections
and focus only on the CL corrections.

There are (at least) two popular ways to parameterise the PMNS matrix; firstly
one can write $U_{\mathrm{PMNS}}=U_{23}U_{13}U_{12}$ with 
\begin{align}
 U_{12}&=\begin{pmatrix}
         c_{12} & s_{12}\exp\left(-i\delta_{12}\right) & 0 \\
	 -s_{12}\exp\left(i\delta_{12}\right) & c_{12} & 0 \\
	 0 & 0 & 1
        \end{pmatrix} , \\
 U_{13}&=\begin{pmatrix}
         c_{13} & 0 & s_{13}\exp\left(-i\delta_{13}\right) \\
	 0 & 1 & 0 \\
	 -s_{13}\exp\left(i\delta_{13}\right) & 0 & c_{13} 
	 \end{pmatrix} ,\\
 U_{23}&=\begin{pmatrix}
         1 & 0 & 0 \\
	 0 & c_{23} & s_{23}\exp\left(-i\delta_{23}\right) \\
	 0 & -s_{23}\exp\left(i\delta_{23}\right) & c_{23}
        \end{pmatrix}.
\end{align}
Here, $c_{ij}$ and $s_{ij}$ stand for $\cos\theta_{ij}$ and $\sin\theta_{ij}$
respectively and the 3 remaining unphysical phases have been rotated away, see
e.g. \cite{King:2005CSD}. Individual rotation matrices $U_{e_L}$ and
$U_{\nu_L}^{\dagger}$ are parameterised in the same way with relevant superscripts. The
second parameterisation is that used by the PDG \cite{Nakamura:2010zzi}, with
a Dirac phase $\delta$ and Majorana phases $\alpha_2$ and $\alpha_3$; this is
constructed as
$U_{\mathrm{PMNS}}^{\mathrm{PDG}}=R_{23}U^{\mathrm{PDG}}_{13}R_{12}P$ where
the $R_{ij}$ are standard orthogonal rotations,
$U_{13}^{\mathrm{PDG}}=U_{13}\left(\delta_{13} = \delta\right)$ and
$P=\mathrm{diag}(1,e^{i\frac{\alpha_2}{2}},e^{i\frac{\alpha_3}{2}} )$. A
comparison of the two parameterisations, after performing a global phase
redefinition to absorb remaining unphysical phases and obtain consistency with
the convention stated in the introduction, 
shows that \cite{King:2005CSD}  
\begin{align}
 \delta=&\;\delta_{13}-\delta_{23}-\delta_{12}, \label{eqn:PDGa}\\  
 \alpha_2=&\;-2\delta_{12},   \label{eqn:PDGb}\\
 \alpha_3=&\;-2\left(\delta_{12}+\delta_{23}\right). \label{eqn:PDGc}
\end{align}

We can now write the parameters of $U_{\mathrm{PMNS}}$ in terms of the
neutrino mixing 
parameters, with perturbative corrections from the charged lepton sector as
follows \cite{King:2005CSD} (neglecting $\theta_{13}^e$ and $\theta_{23}^e$ as
they are small),\footnote{We note that in order to derive these equations
  consistently to first order, the Majorana phases from
  Eqs.~\eqref{eqn:PDGa}-\eqref{eqn:PDGc} must be redefined by a correction of
  order $\theta_{13}^{\nu}$; this is however only a subtlety in the derivation and
  therefore we do not explicitly demonstrate this redefinition, merely point
  it out to the reader.} 
\begin{align}
 s_{23}\exp\left(-i\delta_{23}\right)&\approx
 s_{23}^{\nu}\exp\left(-i\delta_{23}^{\nu}\right)
, \label{eqn:PMNSa}\\ 
 s_{13}\exp\left(-i\delta_{13}\right)&\approx \theta_{13}^{\nu}\exp\left(-i\delta_{13}^{\nu}\right)-\theta_{12}^e s_{23}^{\nu}\exp\left(-i\left(\delta_{23}^{\nu}+\delta_{12}^e\right)\right), \label{eqn:PMNSb}\\ 
 s_{12}\exp\left(-i\delta_{12}\right)&\approx s_{12}^{\nu}\exp\left(-i\delta_{12}^{\nu}\right)-\theta_{12}^e c_{23}^{\nu}c_{12}^{\nu}\exp\left(-i\delta_{12}^e\right). \label{eqn:PMNSc}
\end{align}

The dominance of the first term in Eq.~\eqref{eqn:PMNSc} allows us to approximate
$\delta_{12}\approx\delta_{12}^{\nu}$, while Eq.~\eqref{eqn:PMNSa}
gives directly $\delta_{23}\approx\delta_{23}^{\nu}$. The phase $\delta_{13}$
 requires a more careful treatment, since the first term of Eq.~\eqref{eqn:PMNSb}
is larger but not dominant enough to drop the second term. It turns out to be
possible to write  
\begin{equation}
 \delta_{13}\approx\delta_{13}^{\nu}
-\frac{\theta_{12}^e s_{23}^{\nu}}{\theta_{13}^{\nu}}
\sin\left(\delta_{23}^{\nu}-\delta^\nu_{13}+\delta_{12}^e\right) \ ,
\end{equation}
 assuming that $\frac{\theta^e_{12}s^\nu_{23}}{\theta_{13}^\nu}$ is small.\footnote{With $\theta^e_{12} \sim \frac{\theta_C}{3}$, $s^\nu_{23} \sim 
 \frac{1}{\sqrt{2}}$ and $\theta^\nu_{13} \sim 0.15$ we obtain a numerical
 value of $\frac{\theta^e_{12}s^\nu_{23}}{\theta_{13}^\nu} \sim \frac{1}{3}$.}
Then the physical Dirac oscillation phase can be approximated by
\be
\delta ~\approx ~ \delta^\nu_{13}  -\delta^\nu_{23} -\delta^\nu_{12} -\frac{\theta_{12}^e s_{23}^{\nu}}{\theta_{13}^{\nu}}
\sin\left(\delta_{23}^{\nu}-\delta^\nu_{13}+\delta_{12}^e\right) \ .
 \label{delta}
\ee

Turning to the resulting mixing angles, we first observe that the TM mixing of
the neutrino sector must necessarily be a small deviation from TB mixing. We
can therefore express our results using the {\em neutrino} TB deviation parameters
\cite{King:2007pr},  
\be
 \sin \theta^{\nu}_{12} = \frac{1}{\sqrt{3}}(1+s^{\nu})\ , \quad
\sin \theta^{\nu}_{23}  =  \frac{1}{\sqrt{2}}(1+a^{\nu})\  , \quad 
 \sin \theta^{\nu}_{13} =   \frac{r^{\nu}}{\sqrt{2}}\ ,
\label{rsareminder}
\ee
where here these parameters refer only to the neutrino sector. 
In terms of angles and phases, using Eqs.~\eqref{eqn:PMNSa}-\eqref{eqn:PMNSc}
(see, e.g. \cite{Antusch:2008yc} for a discussion of this procedure), we can
then write the TB deviation parameters for the complete lepton mixing in terms of the TB deviations parameters 
in the neutrino sector and the charged lepton corrections as, 
\begin{align}
 a&\approx 
a^\nu, \label{eqn:perturba}\\ 
 r&\approx
 \modulus{r^{\nu}\exp\left(-i\delta_{13}^{\nu}\right)
 - \theta^e_{12}  \exp\left(-i\left(\delta^{\nu}_{23}
+\delta_{12}^e
\right)\right)}, \label{eqn:perturbb}\\ 
 s&\approx
s^\nu - \theta_{12}^e
  \cos\left(\delta^\nu_{12}-\delta^e_{12}\right)   .
\label{eqn:perturbc}
\end{align}
With the neutrino mixing being of TM form as given in Eq.~\eqref{TM}, the
deviation parameters of the neutrino sector can be shown to satisfy, see
\cite{King:2007pr,King:2010bk,King:2011zj}, $s^{\nu}=0$ and
$a^{\nu}\approx-\frac{r^{\nu}}{2}\cos\delta^{\nu}$. Using
this and the fact that $\theta^e_{12}\sim \frac{\theta_C}{3}$ and Eq.~\eqref{delta}, the above
equations for the tri-bimaximal deviation parameters may be further simplified
to first order as 
\begin{align}
 a&\approx 
-\frac{r^{\nu}}{2}\cos\delta , \label{eqn:perturbsimpa}\\ 
 r&\approx r^{\nu}- \frac{\theta_C}{3}\cos\left(\delta^{\nu}_{23}
-\delta_{13}^{\nu}
+\delta^e_{12}
\right), \label{eqn:perturbsimpb} \\
 s&\approx -\frac{\theta_C}{3}\cos\left(\delta^{\nu}_{12}-\delta^e_{12}\right), \label{eqn:perturbsimpc}
\end{align}
again assuming that $\frac{\theta_{12}^e s^\nu_{23}}{\theta_{13}^\nu} \sim
\frac{\theta_C}{3r^\nu}$ is small. 
In the limit that charged lepton corrections are switched off, the above results 
reduce to the usual TM sum rules \cite{King:2007pr,King:2010bk,King:2011zj}, $s\approx 0$ and
$a\approx-\frac{r}{2}\cos\delta$. In the limit that the neutrino mixing angle $\theta_{13}^{\nu}$ 
is switched off the above results reduce to the usual TB sum rules~\cite{King:2005bj}, $s\approx r \cos \delta$
where $r\approx \theta_C/3$ and $\delta \approx \delta^e_{12} - \delta^{\nu}_{12}$.

The results in Eqs.~\eqref{eqn:perturbsimpa}-\eqref{eqn:perturbsimpc} imply the relatively simple sum rule bounds:
\begin{align}
|s|& \leq \frac{\theta_C}{3}, \label{1}\\ 
|a|& \leq \frac{1}{2} (r+ \frac{\theta_C}{3}) |\cos \delta|, \label{2}
\end{align}
where, again, $r,s,a$ are the tri-bimaximal deviation parameters, 
in particular $r\approx \sqrt{2}\theta_{13}$,
$\delta$ is the CP violating oscillation phase,
and $\theta_C$ is the Cabibbo angle. We emphasise that these bounds do not include 
RG and CN corrections, which however are expected to be rather small for the case of hierarchical
neutrino masses. For example, assuming $\theta_{13}\sim 8^\circ$ we find $r\approx 0.2$, 
and using $\theta_C /3 \approx 0.075$ these bounds become $|s|\leq 0.075$
and $|a|<0.14 |\cos \delta|$. The present approximate limits from the global fit $|a|<0.08$,
$-0.06<s<0$ quoted in Eq.~\eqref{rsafit} are nicely consistent with these sum rule bounds.

\section{Conclusions}
\label{sec:conclusion}
Recent T2K, MINOS and Double CHOOZ results, together with global fits of mixing parameters, 
indicate a sizeable reactor
angle~$\theta_{13}\sim 8^\circ$ which, if confirmed, would rule out TB lepton mixing. 
On the other hand, the TB predictions $\sin \theta_{23} = 1/\sqrt{2}$ and 
$\sin \theta_{12} = 1/\sqrt{3}$ remain in agreement with global fits and continue to provide tantalising 
hints for an underlying Family Symmetry. For example, an
$A_4$ family symmetry model including additional flavons in the ${\bf 1'}$ and
${\bf 1''}$ representations leads to TM
neutrino mixing which maintains the prediction $\sin \theta_{12} \approx1/\sqrt{3}$,
at least approximately, while allowing an arbitrarily large reactor angle.
Indeed, as discussed in a previous paper by two of us, 
the problem in this model is in explaining why 
the reactor angle should be smaller than the atmospheric or solar angles,
which follows from the fact that the additional flavons would be expected to have 
VEVs of the same order as the other TB flavon VEVs, with all undetermined coefficients 
being of order unity. However, apart from this drawback, such a model provides a simple example
of a Family Symmetry model with a non-zero reactor angle.

In this paper we have proposed a SUSY GUT of Flavour with a non-zero $\theta_{13}$
based on $A_4$ Family Symmetry with additional flavons in the ${\bf 1'}$ and
${\bf 1''}$ representations, and an $SU(5)$ GUT group. 
The model involves an additional continuous $U(1)$ family symmetry as well
as three discrete symmetries designed to control the operator structure of the model.
All flavon representations of $A_4$ are populated, and the main flavon content of the quark sector is copied from the neutrino sector. 
The vacuum alignment is obtained using the conventional $F$-term mechanism. 
NLO terms to the mass matrices are negligible, demonstrating 
the stability of the LO matrix textures.
The resulting model exhibits TM mixing in the neutrino sector, with the physical lepton
mixing involving charged lepton corrections, which in turn are related to quark mixing angles.
In particular, the model involves a Georgi-Jarlskog relation, leading to bounds on the TB deviation parameters
$|s|\leq \frac{\theta_C}{3}$, 
$|a|\leq \frac{1}{2} (r+ \frac{\theta_C}{3})|\cos \delta|$, up to RG and CN corrections, which are 
in good agreement with current global fits. The considered model shows that 
it is possible to accommodate $\theta_{13}\sim 8^\circ$, within a SUSY GUT of Favour
which relates quark and lepton masses and mixing angles, while continuing to 
provide an explanation for the TB nature of the solar and atmospheric lepton mixing angles.

\section*{Acknowledgments}
SFK acknowledges partial support 
from the STFC Consolidated ST/J000396/1 and EU ITN grant UNILHC 237920 (Unification in
the LHC era).

\appendix
\makeatletter
\def\@seccntformat#1{Appendix\ \csname the#1\endcsname:\;}
\makeatother

\section{Vacuum alignment}
\label{sec:appenda}
In order that the flavon fields obtain the alignment presented in Eq.~\eqref{eq:flavonalign}, their potential must be minimised in the correct way. We follow the method of \cite{King:2011zj} very closely. The leading order contributions to the driving superpotential are:
\begin{equation}
\begin{split}
 \label{eqn:alignment}
W_{0}&=\varphi_T^0\left(g_1\sigma\varphi_T+g_2\varphi_T\varphi_T\right)+\varphi_S^0\left(g_3\varphi_S\varphi_S+g_4\varphi_S\xi+g_4'\varphi_S\xi'+g_4''\varphi_S\xi''\right) \\
     &+\xi^0\left(g_5\varphi_S\varphi_S+g_6\xi\xi+g_7\xi'\xi''\right).
\end{split}
\end{equation}
Here, $g_1\vev{\sigma}=M$ which appears in the vacuum alignment of
\cite{King:2011zj}; this is required since $\varphi_T$ is charged under the
auxiliary symmetries and so the original structure
$\varphi_T^0\left(M\varphi_T+\varphi_T\varphi_T\right)$ that drives the
$\varphi_T$ alignment cannot be used.
Minimising with respect to $\varphi_T^0$ gives
\begin{equation}
\label{eqn:phit}
\vev{\varphi_T}=v_T\begin{pmatrix}
				1 \\
				0 \\
				0
			  \end{pmatrix},\quad v_T=-\frac{g_1\vev{\sigma}}{2g_2}.
\end{equation}
The conditions from $\varphi_S^0$ and $\xi^0$ are
\begin{equation}
\label{eqn:phis}
2g_3\begin{pmatrix}
         s_1^2-s_2s_3 \\
         s_2^2-s_3s_1 \\
         s_3^2-s_1s_2
        \end{pmatrix}+g_4u\begin{pmatrix}
			          s_1 \\
			          s_3 \\
			          s_2
			         \end{pmatrix}+g_4'u'\begin{pmatrix}
							   s_3 \\
							   s_2 \\
							   s_1
							 \end{pmatrix}+g_4''u''\begin{pmatrix}
										      s_2 \\
										      s_1 \\
										      s_3
										     \end{pmatrix}=\begin{pmatrix}
													0 \\
													0 \\
													0
												          \end{pmatrix},
\end{equation}
\begin{equation}
\label{eqn:xi}
g_5\left(s_1^2+2s_2s_3\right)+g_6u^2+g_7u'u''=0.
\end{equation}
Here, $\vev{\varphi_{S_i}}=s_i$, $\vev{\xi}=u$, $\vev{\xi'}=u'$ and $\vev{\xi''}=u''$. The solutions to these equations are
\begin{equation}
\vev{\varphi_S}=v_S\begin{pmatrix}
			     1 \\
			     1 \\
			     1
			   \end{pmatrix},\quad v_S^2=-\frac{g_6u^2+g_7u'u''}{3g_5},\quad u=-\frac{g_4'u'+g_4''u''}{g_4}
\label{eqn:phisalign}.
\end{equation} 
As in \cite{Altarelli:2005yx}, we assume the undetermined singlets obtain
their VEVs as a result of their soft mass parameters $m^2_s$ (where $s$ stands
for singlet) being driven negative in some portion of parameter space.

\section{Higher order operators}
\label{sec:appendb}
There are many higher order corrections to the mass matrices presented in
Section~\ref{sec:masses} of this paper; most of these give negligible
contributions to masses and mixings. In Tables \ref{tab:higherorder} and
\ref{tab:higherorder2} we give suppressions and example terms of the NLO
operators for each sector; it can be seen that none of these will change the
LO results significantly.

\begin{table}
	\begin{center}\small
		\begin{tabular}{|c|c|c|c|}
			\hline
				Term & Contributes to  & NLO Example & c.f. LO\\ \hline 
				$\left.FNH_{\bf 5}\right._{\vphantom{\big(}}^{\vphantom{\big(}}$& $m_D$  & $\varphi_T ^2 \theta^2\theta'\theta'' \sim \epsilon^6$ & 1 \\ \hline
				$NN^{\vphantom{\big(}}_{\vphantom{\big(}}$ & $M_R$  & $\varphi_T ^2 \theta^2 \left(\theta''\right)^2\xi'' \sim \epsilon^7$ & $\epsilon$ \\ \hline
				 $^{\vphantom{\big(}}$  & $\left(M_d\right)_{11}$  & $\sigma \theta ' \left(\theta ''\right)^4\sim\epsilon^7$ & $\epsilon^6$  \\ 
				$F\varphi_TH_{\overline{\bf 5}}T_1^{\vphantom{\big(}}$ & $\left(M_d\right)_{12}$  & $\sigma  \left(\theta '\right)^2\left(\theta''\right)^3\sim\epsilon^7$ & $\epsilon^4$\\
				 $^{\vphantom{\big(}}$ & $\left.\left(M_d\right)_{13}\right._{\vphantom{\big(}}$  & $\sigma \left(\theta ''\right)^5\sim\epsilon^7$ & $\epsilon^4$ \\ \hline
				 $^{\vphantom{\big(}}$ & $\left(M_d\right)_{11}$  & $\sigma ^2 \theta^5 \theta ''\sim\epsilon^9$ & $\epsilon^6$\\ 
				$F\varphi_TH_{\overline{\bf 45}}T_1^{\vphantom{\big(}}$ & $\left(M_d\right)_{12}$  & $\sigma\theta \left(\theta '\right)^2 \theta ''\sim\epsilon^6$ & $\epsilon^4$ \\ 
				 $^{\vphantom{\big(}}$ & $\left.\left(M_d\right)_{13}\right._{\vphantom{\big(}}$  & $\sigma \theta \left(\theta '\right)^3\sim\epsilon^6$ & $\epsilon^4$\\ \hline
				 $^{\vphantom{\big(}}$ & $\left(M_d\right)_{21}$  & $\sigma \theta^4 \theta'\sim\epsilon^7$ & $\epsilon^4$ \\ 
				$F\varphi_TH_{\overline{\bf 5}}T_2^{\vphantom{\big(}}$ & $\left(M_d\right)_{22}$ & $\left(\theta '\right)^3\sim\epsilon^4$ & $\epsilon^3$  \\ 
				 $^{\vphantom{\big(}}$ & $\left.\left(M_d\right)_{23}\right._{\vphantom{\big(}}$  & $\theta ' \left(\theta ''\right)^2\sim\epsilon^4$ & $\epsilon^3$ \\ \hline
				 $^{\vphantom{\big(}}$ & $\left(M_d\right)_{21}$  & $\sigma^2\theta ^3\left(\theta''\right)^3\sim\epsilon^9$ & $\epsilon^4$ \\ 
				$F\varphi_TH_{\overline{\bf 45}}T_2^{\vphantom{\big(}}$ & $\left(M_d\right)_{22}$  & $\sigma^2\theta ^3\theta'\left(\theta''\right)^2\sim\epsilon^9$ &  $\epsilon^3$\\ 
				 $^{\vphantom{\big(}}$ & $\left.\left(M_d\right)_{23}\right._{\vphantom{\big(}}$  & $\sigma^2\theta ^3\left(\theta'\right)^2\theta''\sim\epsilon^9$ & $\epsilon^3$  \\ \hline
				 $^{\vphantom{\big(}}$ & $\left(M_d\right)_{31}$  & $\sigma ^3 \theta ' \left(\theta ''\right)^5\sim\epsilon^{10}$ & $\epsilon^7$ \\ 
				$F\varphi_TH_{\overline{\bf 5}}T_3^{\vphantom{\big(}}$ & $\left(M_d\right)_{32}$ & $\sigma^2\theta^2\left(\theta''\right)^2\sim\epsilon^7$ &  $\epsilon^6$  \\ 
				 $^{\vphantom{\big(}}$ & $\left.\left(M_d\right)_{33}\right._{\vphantom{\big(}}$ & $\sigma ^2 \theta ^2 \theta ' \theta ''\sim\epsilon^7$ & $\epsilon$ \\ \hline
				 $^{\vphantom{\big(}}$ & $\left(M_d\right)_{31}$  & $\sigma^3\theta\theta'\left(\theta''\right)^3\sim\epsilon^9$ & $\epsilon^7$ \\ 
				$F\varphi_TH_{\overline{\bf 45}}T_3^{\vphantom{\big(}}$ & $\left(M_d\right)_{32}$  & $\sigma ^3 \theta   \left(\theta '\right)^2 \left(\theta ''\right)^2\sim\epsilon^9$ & $\epsilon^6$ \\ 
				 $^{\vphantom{\big(}}$ & $\left.\left(M_d\right)_{33}\right._{\vphantom{\big(}}$ & $\sigma^3\theta\left(\theta''\right)^4\sim\epsilon^9$ & $\epsilon$   \\ \hline
		\end{tabular}\end{center}
	\caption{NLO corrections in the model. The first column shows each
          basic term that exists in the neutrino, down quark (and charged
          lepton) Yukawa superpotential, as specified in the second column. 
          A bunch of flavons is appended to these basic terms to obtain the
          complete term invariant under the symmetries. In the third column we
          give  an example of such a bunch of flavons at NLO and the order of
          its contribution, to be compared to the LO contribution given
          in the final column. Note that in the terms
          contributing to $M_d$, there is a flavon $\varphi_T$ already present
          in the basic term. It is furthermore not specified whether the LO term comes
          from an ${H_{\bf{\overline{5}}}}$ or an ${H_{\bf {\overline{45}}}}$; the reader may refer back to Eq.~\eqref{eqn:down} if required.}
\label{tab:higherorder}
\end{table}

\begin{table}[t]
	\begin{center}\small
		\begin{tabular}{|c|c|c|c|c|}
			\hline
				Term & Contributes to & NLO  Example  & c.f. LO \\ \hline 
				$T_1T_1H_{\bf 5}^{\vphantom{\big(}}$ & $\left.\left(M_u\right)_{11}\right._{\vphantom{\big(}}$ & 
				  $\sigma^2\xi^4\xi''\sim \overline{\epsilon}^7$ & $\overline{\epsilon}^6$  \\ \hline
				$T_1T_2H_{\bf 5}^{\vphantom{\big(}}$ & $\left.\left(M_u\right)_{12},\;\left(M_u\right)_{21}\right._{\vphantom{\big(}}$  & 
				 $\sigma  \theta ^6 \theta ' \theta ''\sim \overline{\epsilon}^9$  & $\overline{\epsilon}^6$\\ \hline
				$T_1T_3H_{\bf 5}^{\vphantom{\big(}}$ & $\left.\left(M_u\right)_{13},\;\left(M_u\right)_{31}\right._{\vphantom{\big(}}$ & 
				  $\sigma \left(\theta '\right)^3 \left(\theta''\right)^2\sim \overline{\epsilon}^6$ & $\overline{\epsilon}^3$ \\ \hline
				$T_2T_2H_{\bf 5}^{\vphantom{\big(}}$ & $\left.\left(M_u\right)_{22}\right._{\vphantom{\big(}}$ & 
				$\left(\theta '\right)^5 \theta ''\sim \overline{\epsilon}^6$   & $\overline{\epsilon}^3$ \\ \hline
				$T_2T_3H_{\bf 5}^{\vphantom{\big(}}$ & $\left.\left(M_u\right)_{23},\;\left(M_u\right)_{32}\right._{\vphantom{\big(}}$ & 
				 $\sigma \theta^4  \theta ''\sim \overline{\epsilon}^6$  & $\overline{\epsilon}^3$ \\ \hline
				$T_3T_3H_{\bf 5}^{\vphantom{\big(}}$ & $\left.\left(M_u\right)_{33}\right._{\vphantom{\big(}}$  & 
				  $\sigma ^2 \theta^2 \theta' \theta ''\sim \overline{\epsilon}^6$& $1$ \\ \hline
				$\left.\varphi_T^0\right.^{\vphantom{\big(}}_{\vphantom{\big(}}$ & $W_{0}$  & 
				 $\sigma ^3 \varphi_T \theta^2 \theta' \theta ''\sim \widetilde{\epsilon}^8$ & $\widetilde{\epsilon}^2$ \\ \hline
				$\left.\varphi_S^0\right.^{\vphantom{\big(}}_{\vphantom{\big(}}$ & $W_{0}$  & 
				 $\sigma^2 \varphi_S \theta^2 \left(\theta ''\right)^2 \xi''\sim \widetilde{\epsilon}^8$  & $\widetilde{\epsilon}^2$ \\ \hline
				$\left.\xi^0\right.^{\vphantom{\big(}}_{\vphantom{\big(}}$ & $W_{0}$ & 
				  $\sigma ^2 \theta^2 \left(\theta ''\right)^2 \xi \xi''\sim \widetilde{\epsilon}^8$  & $\widetilde{\epsilon}^2$  \\ \hline
\end{tabular}\end{center}
	\caption{NLO corrections in the model. The first column shows each
          basic term that exists in the up quark Yukawa and vacuum alignment
          sectors, as specified in the second column. A bunch of flavons is
          appended to these basic terms to obtain the complete term invariant
          under the symmetries. In the third column we give  an example of
          such a bunch of flavons at NLO and the order of its contribution, to
          be compared to the LO contribution given in the final column.
	  The notation $\widetilde{\epsilon}$ is simply used
          to denote that we are in a different sector to $\epsilon$ or~$\overline{\epsilon}$.} 
\label{tab:higherorder2}
\end{table}

\pagebreak



\end{document}